# Unveiling the outstanding oxygen mass transport properties of Mn-rich perovskites in grain boundary-dominated La$_{0.8}$Sr$_{0.2}$(Mn$_{1-x}$Co$_x$)$_{0.85}$O$_{3\pm\delta}$ nanostructures

Aruppukottai M. Saranya[1], Alex Morata[1], Dolors Pla[1,2], Mónica Burriel[1,2,3], Francesco Chiabrera[1], Iñigo Garbayo[1], Aitor Hornés[1], John A. Kilner[3,4], Albert Tarancón[1,5]*

[1]Department of Advanced Materials for Energy Applications, Catalonia Institute for Energy Research (IREC), Jardins de les Dones de Negre 1, 08930 Sant Adrià del Besòs, Barcelona, Spain

[2]Univ. Grenoble Alpes, CNRS, LMGP, F-38016 Grenoble, France

[3]Department of Materials, Imperial College London, London, SW7 2AZ, UK

[4]Electrochemical Energy Conversion Division, International Institute for Carbon-Neutral Energy Research (I2CNER), Motooka 744, Nishi-Ku, Fukuoka 819-0395, Japan

[5]ICREA, Passeig Lluís Companys 23, 08010 Barcelona, Spain

**ABSTRACT:** Ion transport in solid-state devices is of great interest for current and future energy and information technologies. A superior enhancement of several orders of magnitude of the oxygen diffusivity has been recently reported for grain boundaries in lanthanum strontium manganites. However, the significance and extent of this unique phenomenon is not yet established. Here, we fabricate a thin film continuous composition map of the La$_{0.8}$Sr$_{0.2}$(Mn$_{1-x}$Co$_x$)$_{0.85}$O$_{3\pm\delta}$ family revealing a substantial enhancement of the grain boundary oxygen mass transport properties for the entire range of compositions. Through isotope-exchange depth profiling coupled to secondary ion mass spectroscopy, we show that this excellent performance is not directly linked to the bulk of the material but to the intrinsic nature of the grain boundary. In particular, the great increase of the oxygen diffusion in Mn-rich compositions unveils an unprecedented catalytic performance in the field of Mixed Ionic Electronic Conductors. These results present grain boundaries engineering as a novel strategy for designing highly performing materials for solid state ionics based devices.

## INTRODUCTION

Advanced energy conversion and storage devices, such as batteries, solar-to-fuel converters, fuel cells and supercapacitors as well as novel solid state electrochemical information storage and logic devices such as resistive switching memories are based on Mixed Ionic-Electronic Conductors (MIECs).[1] Recently, an intensive effort has been focused on looking for the enhancement of the ionic transport in MIECs.[2] Among other strategies, the implementation of nanoionics concepts based on engineering charge and mass transport properties at the nanoscale have been proposed to tune the functionality of interface-dominated structures like nanostructured or multilayered thin-film MIECs.[2] The search for enhanced ionic conductivity through the design of oxide interfaces has been pursued since Sata *et al.* showed superior anionic conductivity in the CaF$_2$/BaF$_2$ heterostructures due to space charge effects.[3] For oxygen ion conductors, multilayering epitaxial thin films have also been successfully proposed to tune the lattice strain subsequently increasing the ionic conductivity.[4] However, the implementation of this type of heterostructures still remains a challenge since it is limited to a certain number of substrates and to a demanding lateral architecture. Therefore, clear advantages would arise if this enhancement of mass and charge transport properties could occur at intrinsic material interfaces, such as grain boundaries (GBs).[5] This will open new technological perspectives for GB-dominated MIEC materials in advanced devices such as micro-solid oxide fuel cells,[2,6,7] lithium ion batteries[8,9] or nanoionics-based resistive switching devices.[10,11]

Recently, independent works from Navickas *et al.*,[12,13] Usiskin *et al.*[15] and Tarancón and co-workers[14,16] have shown a great enhancement of the mass transport properties of nanostructures of La$_{0.8}$Sr$_{0.2}$MnO$_{3+\delta}$ (LSM). Several orders of magnitude of enhancement were reported for the oxygen mass transport compared to the bulk properties in highly dense polycrystalline LSM thin films with vertically aligned grain boundaries. This enhancement was mostly attributed to the high concentration of strain-induced defects at the grain boundary level that even allowed the transformation of a mainly electronic conductor like LSM into a good mixed ionic electronic conductor comparable to the state-of-the-art La$_{0.8}$Sr$_{0.2}$CoO$_{3-\delta}$ (LSC). In this sense, grain boundary engineering can be considered a novel approach for substantially modifying the properties of oxygen ionic conductors and MIECs. However, a generalization of the enhancement of mass transport properties of MIEC oxides at the grain boundary level is far from being fully established. For instance, recent works by De Souza and co.[17,18] and Yildiz and co.[19,20] showed a deleterious effect on the ionic diffusivity with the presence of dislocations in the grain boundaries for SrTiO$_3$ and CeO$_2$. Therefore, to fully understand the oxygen diffusion enhancement in MIECs further experimental and theoretical work is necessary, to clarify and complement currently existing data. This will allow elucidating the origin of the phenomenon and controlling and tuning the mechanisms taking place in these materials.

In order to evaluate the scope of the grain boundary engineering approach, it is of particular interest to understand if the enhancement occurring in materials with low ionic conductivity is also happening in materials showing reasonably good mass transport properties at the bulk level. This can be rationally done by studying the effect of a high density of grain boundaries in the mass transport properties of a material system with tuneable oxygen diffusivity and surface exchange. In this direction, it is well known that the substitution of Mn by Co in the $La_{0.8}Sr_{0.2}Mn_{1-x}Co_xO_{3\pm\delta}$ (LSMC) perovskite promotes significant changes in the oxygen mass transport properties converting a mainly electronic conductor such as the oxygen hyper-stoichiometric LSM into a reasonably good MIEC, like the oxygen deficient perovskite LSC.[21,22]

In this work, we fabricate a thin film pseudo-binary system $La_{0.8}Sr_{0.2}(Mn_{1-x}Co_x)_{0.85}O_{3\pm\delta}$ ($x \approx 0$ to 1) by using combinatorial Pulsed Laser Deposition (C-PLD).[23–31] After proving the high quality of the films with fully dense and columnar-type nanostructures, required for accurate oxygen diffusion measurements, we show the evolution of the mass transport properties with cobalt content as directly measured by $^{18}O$ Isotope Exchange Depth Profiling coupled to Time of Flight Secondary Ion Mass Spectroscopy (IEDP- ToF-SIMS). Numerical simulations of the obtained diffusion profiles allow us to quantify the relevant bulk and grain boundary oxygen diffusion and surface exchange coefficients for different temperatures and cobalt contents. By comparing both sets of mass transport coefficients, we clearly prove an enhancement of grain boundary properties independently of the cobalt content. Finally, the electronic conduction properties of the LSMC family are measured, showing no direct correlation between the electronic properties and oxygen incorporation and therefore linking the origin of the enhanced properties to the enabling role of oxygen vacancies present at the grain boundary level.

**EXPERIMENTAL SECTION**

Small area, large area and combinatorial depositions of LSM, LSC, yttria-stabilized zirconia YSZ (buffer layer) and LSMC thin film were carried out by PLD. A PLD-5000 system from PVD products was used, coupled to a Lambda Physik COMPex PRO 205 KrF excimer Laser (wavelength $\lambda = 248$ nm, pulse duration 20 ns, max Power $P = 30$ W, max repetition rate $f = 50$ Hz). Commercial targets with nominal compositions $La_{0.8}Sr_{0.2}MnO_3$ (LSM), $La_{0.8}Sr_{0.2}CoO_3$ (LSC), (purity $\approx 99.95\%$, density $\rho = 99.99$ %, diameter $D = 2$ inch, thickness $t = 4$ mm) from SurfaceNet GmBH Germany and home-made target 8 mol-% $Y_2O_3$ stabilized $ZrO_2$ (8YSZ) (density $\rho = 99.99$ %, diameter $D = 4$ inch, thickness $t = 5$ mm) were used as targets in PLD for thin film fabrication.

Structural X-Ray Diffraction (XRD) studies were performed using a Bruker D8 diffractometer equipped with CuKα radiation ($\lambda = 1.54184$ Å). The measurements were carried out in $\theta - 2\theta$ (offset) scanning mode with step size 0.02 °. The microstructure of LSM, LSC, YSZ and LSMC thin films were studied by Scanning Electron Microscopy (SEM) in a ZEISS AURIGA (equipped with Energy-Dispersive Spectroscopy, EDS) microscope. The atomic percentage of the constituent elements in LSM, LSC thin films and LSMC thin film system was studied by EDS and Wavelength-Dispersive Spectroscopy (WDS, model- Jeol JXA-8230) techniques.

Single crystal (100)-oriented 4 inch silicon wafers were employed as substrates for the oxygen exchange experiments. The Si/YSZ/LSMC heterostructures (see further details on the combinatorial sample deposition in the Supporting Information file, **section S.1**) employed for the IEDP measurements were diced into 10 x 10 mm$^2$ by using an automatic dicing saw machine. Chemical cleaning of the sample surface was avoided to preserve, as much as possible, the surface from modification/contamination that could alter the surface exchange properties. A first thermal annealing at the selected exchange temperature in 200 mbar of dry pure (99.9996%) oxygen of normal isotopic abundance for at least thirty times the duration of the isotope exchange annealing was carried out to equilibrate the oxygen stoichiometry of the PLD-deposited thin films. After this equilibration step, the sample was rapidly cooled down, pure oxygen was evacuated from the chamber and labelled oxygen was introduced instead. The isotope exchange was carried out in 200 mbar of dry $^{18}O$ (55.3%) at the exchange temperatures (*i.e.* 600 and 700ºC for x = 0.6 – 0.8 and x < 0.6, respectively) for the selected time (1250 – 1500 s). Finally, the samples were quenched again to room temperature. The exchanged samples were analyzed by ToF-SIMS using a system from IONTOF (TOF.SIMS 5) equipped with a Bismuth Liquid Metal Ion Source (LMIS) pulsed gun incident at 45°.

During the analysis, a 25 keV Bi$^+$ primary ion beam was used to generate the secondary ions via the low mass resolution or burst alignment mode (eight pulses) for analysis and a 2 keV Cs$^+$ beam incident at 45° for sputtering. The secondary ions were measured in the centre of the sputtered crater in areas varying between 50 x 50 and 100 x 100 μm$^2$. The charge compensation during the analysis was solved by flooding low energy electrons from the electron gun.

Electrical measurements were performed in a temperature-controlled Linkam probe station, after cutting the combinatorial sample in 1 x 0.5 cm$^2$ chips. Silver electrodes were used as contacts in 4-points in-plane electrical measurements (see details in Supplementary **section S.5.1**). The LSC parent compound thin film measured was deposited on a sapphire (0001) substrate to avoid crack formation and serve as a reference for the analysis.

**RESULTS AND DISCUSSION**

**Continuous compositional-spread LSMC system**

In order to study the dominating transport properties in nanostructured $La_{0.8}Sr_{0.2}(Mn_{1-x}Co_x)_{0.85}O_{3\pm\delta}$ ($x = 0 - 0.8$), a real continuous compositional-spread LSMC system was deposited in thin film form by combinatorial Pulsed Laser Deposition. PLD is one of the methods of choice for preparing complex oxides thin films due to possible control of the stoichiometric transfer of the target composition and fully dense columnar-type nanostructures suitable for diffusion studies. The employed combinatorial PLD approach is based on the superposition of large-area films of the parent compounds, *i.e.* LSM and LSC, which follow a Gaussian shape (see **sections S.2.1** and **S.2.2** of the Supporting Information file), centred at opposite sides of a 4" wafer substrate (further details on the superposition can be found in **section S.2.3** of the Supporting). The different thickness of the LSM and LSC layers along the wafer defines the relative Co:Mn content, *i.e.* the composition of the final LSMC layer at each position. This *in-situ* synthesis of the LSMC compound by inter-diffusion of Mn and Co cations was achieved by depositing very thin ($t_{max}$ = 1 nm) individual alternate LSM and LSC layers at high substrate temperatures ($T$ = 700ºC, oxygen partial pressure of p = 2.66 Pa)[32–34]. Under these conditions, the calculated effective inter-diffusion time for Co and Mn is ~ 1 s ensuring instantaneous single phase formation. The final LSMC combinatorial sample covering the whole range of compositions ($La_{0.8}Sr_{0.2}(Mn_{1-x}Co_x)_{0.85}O_{3\pm\delta}$; $x \approx 0$ to 1) was grown on a 4-inch (100)-oriented Si wafer covered with a 100-nm thick yttria-stabilized zirconia (YSZ) layer. A reasonable thickness of the LSMC layer (from 80 to 175 nm) and a low Co-content gradient was defined to allow us obtaining a large spread of cobalt intermediate compositions with crack-free microstructures.

A comprehensive structural and microstructural characterization of the combinatorial sample was carried out by various experimental techniques. The structural investigation was based on X-Ray Diffraction (XRD), the morphology was determined by Scanning Electron Microscopy (SEM) and Atomic Force Microscopy (AFM) of the surface, the thickness and composition distributions by SEM-Energy Dispersive Spectroscopy (EDS) and Wavelength Dispersive Spectroscopy (WDS), and the quality of the LSMC/YSZ interface by ToF-SIMS.

The series of XRD patterns acquired along the central axis of the LSMC combinatorial sample (corresponding to a predicted increment of the Co concentration from $x$ = 0.03 to $x$ = 0.94) show single phase deposition of polycrystalline YSZ and LSMC layers (**Figure 1a**). A shift of the perovskite peaks to higher angles was observed by increasing the Co doping concentration. This trend is in agreement with the shrinkage of the unit cell volume due to a continuous decrease of the lattice parameter as a function of the Co-content from approximately $a_{pc}$=3.90 Å to 3.84 Å (where $a_{pc}$ refers to the pseudo-cubic lattice).[35–37] This decrease is generally ascribed to the smaller average ionic radius of $Co^{3+}$ (0.61 Å) compared to $Mn^{3+}$ (0.645 Å) (**Figure S13, section S.3.1**). A change in peak intensity is observed when increasing the Co content (increase of the l00 reflections and decrease of the 111 and 211 reflections). This variation might be related to a preferential orientation appearing when increasing the Co content.

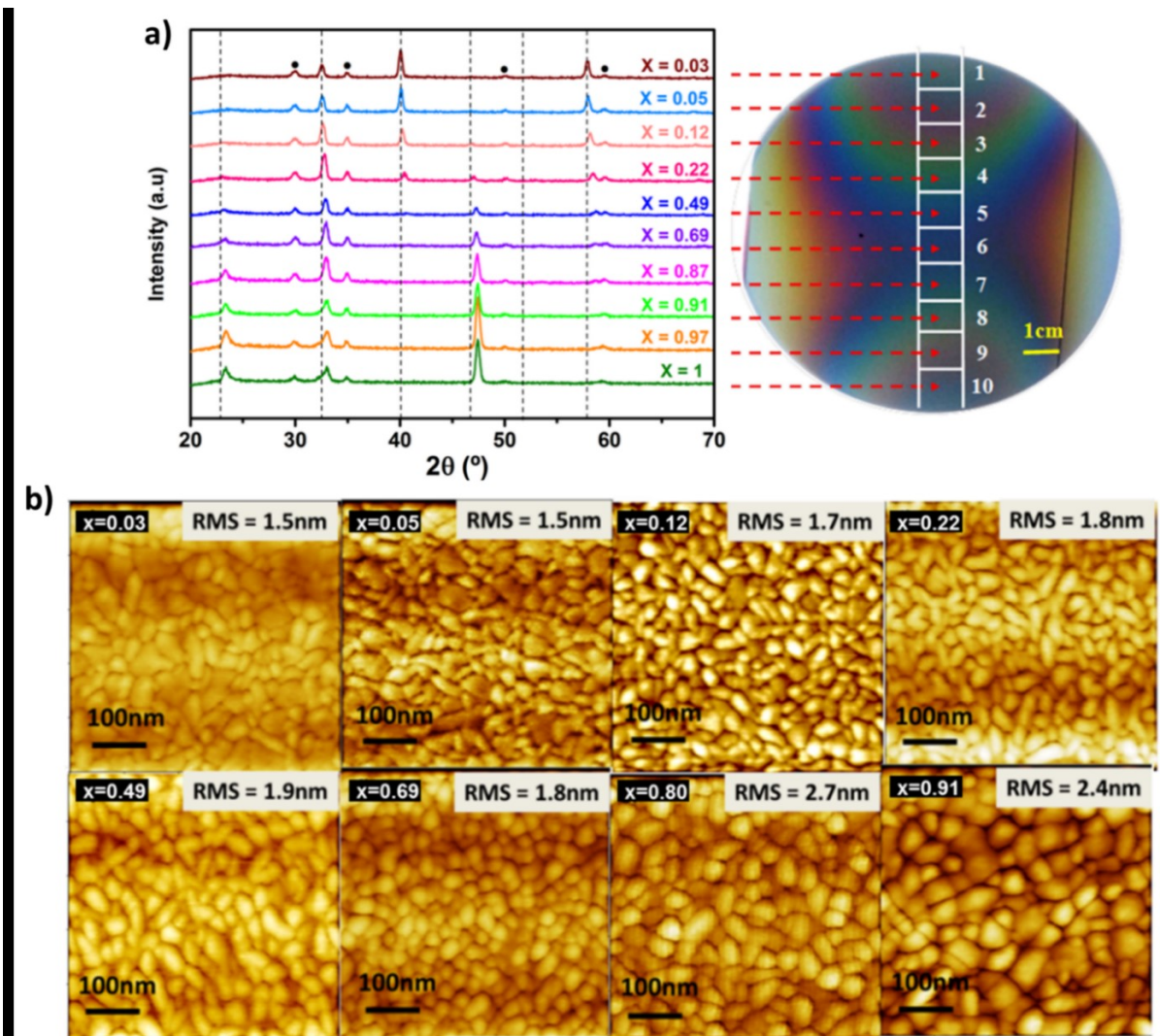

**Figure 1.** (a) XRD patterns of the LSMC system as a function of the Co-content. The arrow marks indicate to which section of the combinatorial sample each XRD pattern corresponds. The vertical dashed lines corresponds to the position of the pseudo-cubic diffraction peaks of bulk LSM ($a_{pc,LSM}$ = 3.89 Å). (b) AFM images of the surface of the LSMC layer for different cobalt concentrations. The increasing evolution of the grain size and surface roughness was analysed as a function of the Co content, obtaining values from 20 to 50 nm and from 1.5 to 2.7 nm, respectively.

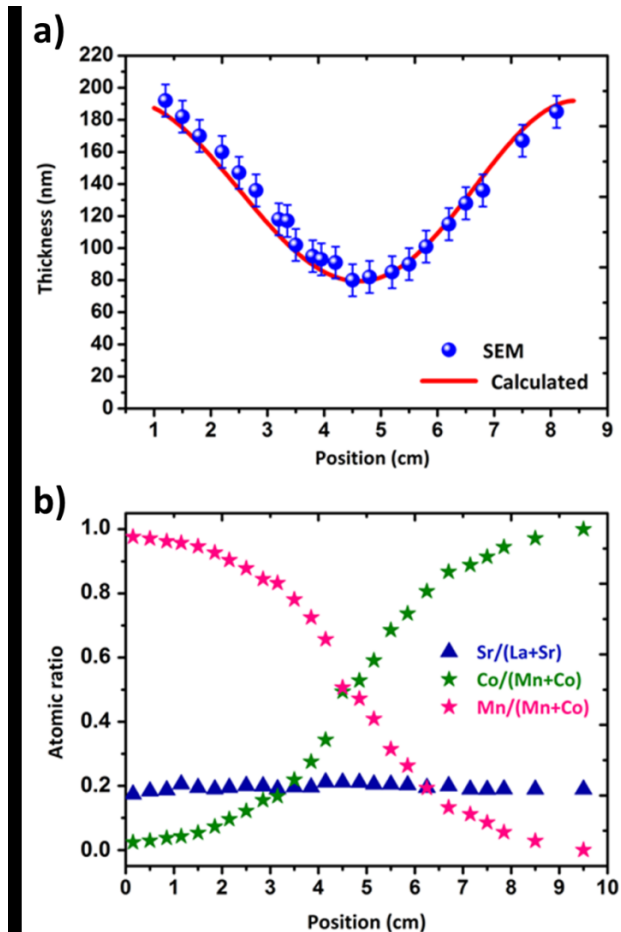

**Figure 2. (a)** Thickness and **(b)** composition along the central axis of the combinatorial sample obtained by SEM and WDS techniques, respectively. The calculated thickness in **(a)** refers to the superposition of the thickness profiles obtained for the individual parent compound layers, as shown in **section S.2.2 and S.2.3** of the Supporting Information file. The Sr/(La+Sr), Co/(Mn+Co) and Mn/(Mn+Co) atomic ratio are included in **(b)**.

Cross-section and top view SEM (**Figure S14 and S15, section S.3.2**) and AFM images (**Figure 1.b**) demonstrate a full dense columnar-type microstructure with grain sizes in the nanoscale (20 – 50 nm) for the LSMC film. Low-magnification SEM surface images show crack-formation for Co-rich compositions with x > 0.8 (**Figure S15**), limiting the current diffusion study to the range of compositions with x < 0.8. The series of AFM and SEM images confirms a morphological evolution with Co-content with an increase in grain size from 20 to 50 nm (**Figure S16, section S.3.3**) and in surface roughness from 1.5 to 2.7 nm (**Figure 1b**).

The thickness distribution along the central axis of the LSMC combinatorial sample was studied by SEM (**Figure 2a**) showing good agreement with the thickness distribution predicted from the superposition of the two parent layers (details in **sections S.2.2 and S.2.3** of the Supporting Information file). Additionally, the composition distribution along the central axis of the combinatorial sample was studied by EDS and WDS. In **Figure 2b**, the atomic ratio of A- and B-site atoms of the LSMC system calculated from WDS is plotted against the wafer position. The atomic ratio of A-site cations (La and Sr) closely follows the nominal content of 80:20 for La:Sr. The atomic B-site ratio of Mn and Co cations follows a Gaussian distribution along the central axis (similarly to the thickness) showing a good agreement with the predicted values (**Figure S17, section S.3.4**). The atomic ratios between La/Sr (A-site), Mn/(Co+Mn) (B-site), (La+Sr)/(Mn+Co) (A/B-site) atoms are also estimated from the atomic concentration, arising an average value of A/B≈ 1.2, independent of the Mn/Co ratio or, in other words, of the position in the wafer. Interestingly, this indicates the presence of a B-site deficient perovskite previously observed for low oxygen partial pressure PLD-deposited LSM layers.[38,39]

The quality of the multilayers was evaluated by ToF-SIMS. A typical SIMS depth profile of the different species contained in the LSMC/YSZ/Si system is presented in **Figure S18, section S.3.5** of the Supporting Information. The high quality of the multilayer is clearly visible by the sharpness of the interfaces, negligible interdiffusion between LSMC and YSZ and the very small Si diffusion in the LSMC layer.

### Oxygen mass transport properties of the LSMC system

The oxygen transport in LSMC thin films was examined by mapping the combinatorial sample using IEDP-SIMS.[40,41] This is a direct method used to determine the oxygen kinetic parameters where oxygen mass transport takes place under zero electrochemical driving force.[42] The parameters acquired from IEDP-SIMS are typically the oxygen self-diffusion coefficient ($D^*$) and the oxygen surface exchange coefficient ($k^*$). In order to determine these oxygen transport parameters for the LSMC system, IEDP-SIMS experiments were carried out at 600 and 700 ºC. Silicon wafer substrates were employed to avoid oxygen diffusion intake from the backside during the exchange process, therefore limiting the incorporation of oxygen to the top surface area. The sputter-time was converted into depth using thickness values of the films measured by SEM. The $^{18}O$ and $^{16}O$ secondary ions measured allowed calculating the $^{18}O$ isotopic fraction as a function of the distance to the top surface, $C'(x)$, corrected for the background isotope fraction ($C'_{bckg}$) and normalized to the $^{18}O$ concentration of the exchange gas ($C'_g$).

The obtained set of normalized isotopic fraction profiles for the whole range of compositions studied for 700 ºC and 600 ºC are presented in **Figure 3a** and **3b**, respectively. Since similar exchange times (1250-1500 s) were employed for all experiments, higher isotopic concentrations and flatter diffusion profiles are roughly correlated to better oxygen mass transport properties. Accordingly, it is clear that increasing the cobalt content and temperature leads to an enhancement of the oxygen diffusion and exchange. At 700 °C the samples with high Co fractions above 0.6 were fully saturated in $^{18}O$ isotope, therefore, the samples corresponding to the Co-rich range (between $x$ =0.6 and 0.8) were analysed at 600 ºC. Note here that a dip in $^{18}O$ concentration was found near the surface for certain compositions, which can be related to back-exchange and/or slight cationic rearrangements near the surface (never affecting the data analysis presented here, see **section S.4.2** for further details).

Similarly to previously reported by the authors for LSM thin films,[14] the shape of the oxygen diffusion profiles indicates that LSMC is not a homogeneous medium (the profiles present deep-penetrating diffusion tails and steps at the LSMC/YSZ interface). These profiles are fully compatible with the existence of narrow fast oxygen diffusion pathways across the LSMC film probably associated to grain boundaries of the columnar nanostructure generated by PLD. Since analytical solutions are not available for diffusion in such a heterogeneous bilayer thin film, the oxygen isotope diffusion profiles of the LSMC/YSZ bilayer were studied by numerical Finite Element Method (FEM) to extract the oxygen transport parameters along the bulk and grain boundaries of the LSMC layer, i.e. $D_b^*, D_{gb}^*$ and $k_b^*, k_{gb}^*$ (YSZ, $D^* = 3 \cdot 10^{-8} cm^2 s^{-1}$,[52] can be considered an oxygen sink in the range of temperatures and thicknesses under study). The geometry of the FEM model consists of a columnar cell made of four-half grains separated by 1-nm thick grain boundaries (see **Figure S19, section S.4.1** for further details of the physical model). The average grain size values and thickness measured by SEM and AFM were used for the construction of the different geometries for the different compositions. Previously reported values of the bulk coefficients for the whole range of cobalt content[21,22] were employed as a starting point for the adjustment. For the grain boundary, manually adjusted values of diffusion and exchange coefficients ($D_{gb}^*$ and $k_{gb}^*$) were used to simulate the experimental profiles. The shape of the experimental profiles was found to be very sensitive to $D_b^*, D_{gb}^*$ and $k_b^*, k_{gb}^*$ values and they were adjusted until a good fit was obtained. As an example, the oxygen isotope profile simulated for a LSMC/YSZ bilayer with 6% Co ($x$ = 0.06) is given in **Figure 3c**. The simulated profiles are in good agreement with the experimental profiles confirming that the FEM model could fit the different regions of the experimental profile. The complete set of oxygen tracer diffusivities and surface exchange coefficients for bulk and GBs of the nanostructured LSMC was simulated and fitted. The 3D isotopic concentration maps generated by simulation after fitting the mass transport parameters clearly support the existence of an oxygen diffusion highway through the LSMC GBs from the top surface to the YSZ (see inset in **Figure 3c**). It is interesting to mention that increasing the Co-content makes the effect of the grain boundary less evident due to a homogenization of the isotopic concentration caused by an increase of the bulk diffusivity, which allows lateral diffusion from the GB into the bulk (**Figure 4a**).

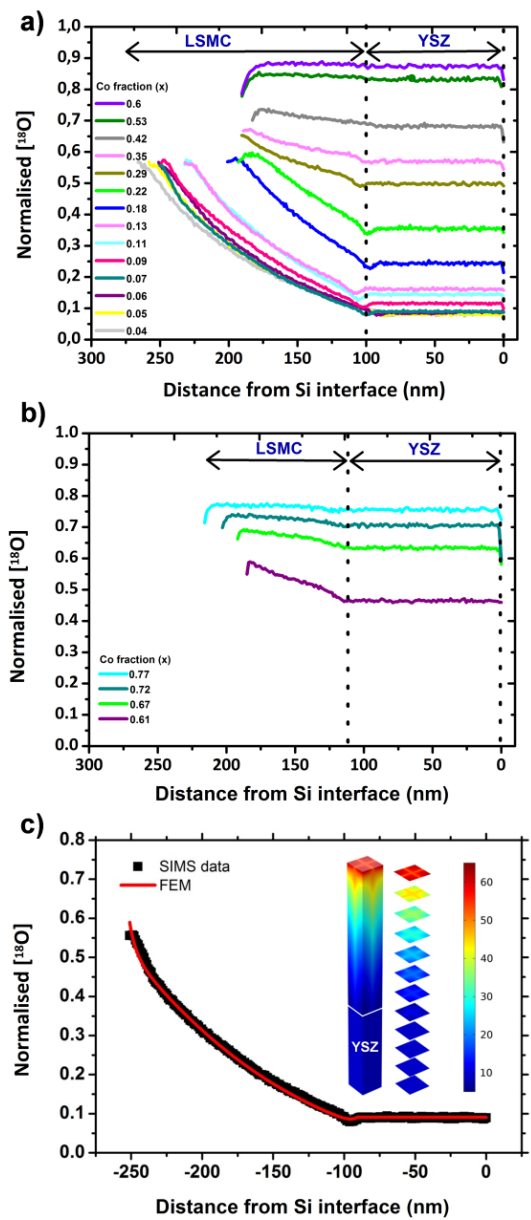

**Figure 3.** Normalized $^{18}$O isotopic fraction depth profile obtained in LSMC/YSZ bilayers deposited on silicon substrates as obtained by IEDP-ToF-SIMS measurement. The x-axis represents the distance from the YSZ/Si interface. Isotope exchange temperatures of (a) 700 °C and (b) 600 °C are represented for different cobalt fractions. (c) Experimental and simulated oxygen isotope concentration profiles corresponding to the sample with 29% Co. The inset shows the cross-section 3D images of the oxygen isotope ($^{18}$O) concentration distribution map obtained by FEM for the same sample.

**Figure 4b and c** show the set of oxygen mass transport coefficients obtained by FEM analysis as a function of the cobalt content and temperature. The values obtained for the bulk coefficients ($D_b^*$ and $k_b^*$) are consistent with literature values[21,22] (also included in **Figure 4b and c** for comparison). On the other hand, for the whole range of cobalt under study, the GB coefficients are orders of magnitude larger than the bulk ones indicating the presence of high diffusion pathways and excellent oxygen surface exchange. This enhancement represents a natural extension of the grain boundary effect previously observed for LSM to the LSMC family.[14,16]

Both grain boundary and bulk diffusivity increase with Co-content (**Figure 4b and 4c**). The origin of the increase in the bulk diffusivity, more noticeable for x > 0.2, is due to the already reported increase of the oxygen vacancy ($V_Ö$) concentration for the transition from an oxygen hyper-stoichiometric material (LSM) to a sub-stoichiometric material (LSC).[24,36] Increasing the cobalt content shows a smaller effect on the GB diffusivity, probably due to the high concentration of oxygen vacancies present at the grain boundary level even before the B-site substitution with Co. This was previously suggested by Chiabrera et al.[16] after observing a negligible change of the ionic conductivity of nanostructured LSM when significantly decreasing the oxygen partial pressure.

**Figure 4c** shows an increase of the bulk surface exchange coefficient as a function of the cobalt content that is fully compatible with previously reported values in the literature.[21,22] Complementarily, a very large enhancement of the GB surface exchange coefficient is observed for the whole range of Co content, suggesting that GBs control the oxygen incorporation in this type of interface-dominated LSMC materials. Importantly, $k^*_{gb}$ happens to be roughly constant along the whole set of compositions, and presents values eventually higher than those found in pure LSC for the entire LSMC family.

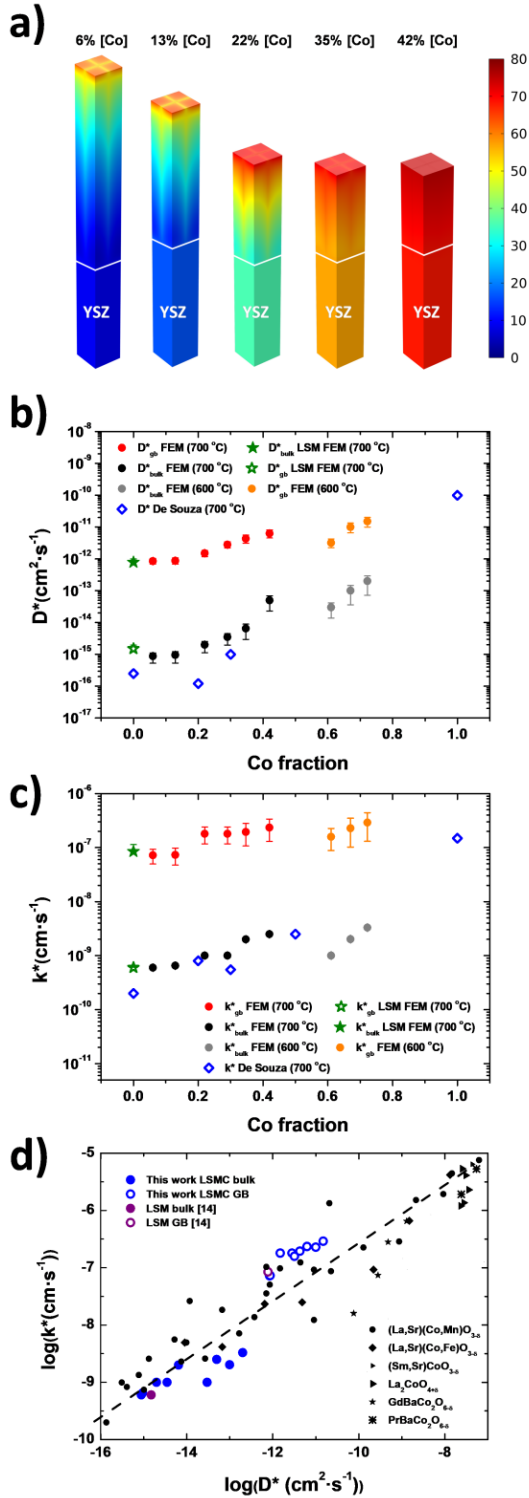

**Figure 4.** (a) Cross-section 3D images of the oxygen isotope ($^{18}O$) concentration distribution maps obtained by FEM for the whole range of Co content at T=700ºC; (b) Oxygen self-exchange diffusion and (c) surface exchange coefficient along bulk and grain boundary extracted from simulation for T=600 °C and 700 °C. The values are compared with bulk literature values[21,22] and with our previous work on LSM thin films.[15] (d) Correlation of the oxygen surface exchange coefficient $k^*$ with the oxygen self-exchange diffusion coefficient $D^*$ (literature values taken from De Souza et al.[22], Saranya et al.[14] and Tarancón et al.[43]).

Despite the significant increase in mass transport properties of these GBs, the obtained results are still in good agreement with the empirical law, firstly proposed by Kilner and co-workers[22,44] which correlates $D^*$ and $k^*$ in a variety of MIEC perovskites. **Figure 4d** represents this correlation ($\log D^*$ vs. $\log k^*$) for a set of perovskite materials with mixed ionic-electronic conduction reported in the literature, in comparison to the authors' results for interface-dominated LSMC and LSM layers. Notably, the superior properties of the GBs keep the trend reported for single phase perovskites despite the highly defective nature of these interfaces. This directly points out to a similar oxygen incorporation mechanism than the bulk material, although greatly enhanced by the high oxygen vacancy concentration at the grain boundary level.

**Evolution of the electrical properties along the LSMC system**

To gain further insights on the oxygen exchange variation with the Co content, we also analysed the evolution of the in-plane electronic conductivity in the combinatorial sample as a function of the temperature. **Figure 5a** shows the evolution of the conductivity with the inverse temperature for the set of compositions. First, the adiabatic small polaron hopping model (see **section S5** of the Supporting Information) could be fitted for the whole range of Co content, as reported for bulk samples.[35,45,46] The conductivity shows a minimum for intermediate composition (x ~ 0.3), and increases when approaching both parent compounds. Opposite, the associated activation energies present a maximum at the same Co-content (x ~ 0.3, see **Figure 5b**). In general, this critical point is linked, in solid solutions, to the activation of the conduction pathway through the dopant-cation sub-lattice for compositions above the percolation threshold, e.g. x ~ 0.31 in simple cubic lattices. This trend is also in good agreement with the existing literature for bulk LSMC, as presented in the same figure for direct comparison.[35] For bulk LSMC, the substitution of Mn by Co in $La_{1-x}Sr_xMnO_{3+\delta}$ progressively deteriorates the electrical conductivity, because the difference in cations size and electronegativity generate lattice distortions and increase of carrier localization effects.[46,47] Similarly, Mn substitution of Co in pure LSC also reduces the electronic conduction due to the hole trapping effect of Mn cations.[48] Our films follow a similar trend of the activation energy presenting the maximum at x ~ 0.3, likely corresponding to the change in the electronic conduction from the Mn to the Co sub-lattice (also supported by the increase of the pre-exponential factor shown in **Figure S24**). However, our Mn-rich films have higher activation energies than the bulk. This might be related to chemical and structural defects present at the GB, which can lead to modifications of the Mn(Co)-O bonds and therefore to drastic changes of the electronic properties. The bulk behaviour is fully recovered above the percolation threshold for Co-rich films.

This transition of the electronic conduction from the Mn- to the Co-sub-lattice has however not a direct impact on the evolution of the surface exchange coefficient, suggesting that the electronic structure is not limiting the oxygen incorporation. In this direction, recent DFT studies[49] on LSM have shown that the availability of oxygen vacancies is controlling the surface exchange properties; moreover, Liu and co-workers showed that oxygen adsorption is highly favourable in LSM, eventually better than in LSC compounds.[50,51] Here, we experimentally unveil this theoretically predicted behaviour linking it to the high availability of oxygen vacancies at the GB level, which completely change the bulk scenario in LSM for which the performance is limited by the extremely low concentration of oxygen vacancies. Under these conditions, the superior properties of GBs in Mn-rich LSMC compounds allows attaining high oxygen exchange coefficients comparable to those found in pure LSC.

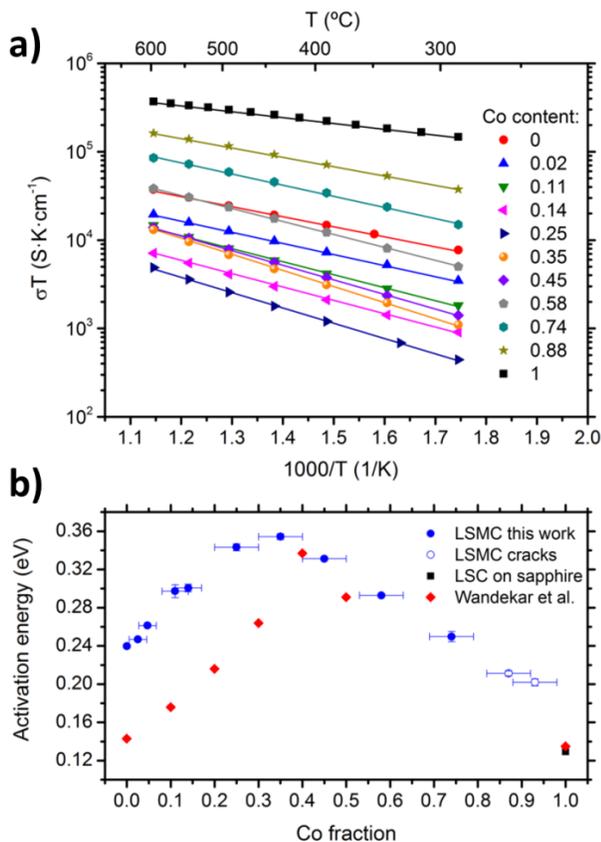

**Figure 5.** (a) In-plane electrical conductivity as a function of temperature for different Co content (b) Activation energy derived from the conductivity measurements as a function of Co content. Lateral errors account for the predictable compositional variations in the LSMC films between the measuring electrodes. The open symbols refer to samples with visible cracks that could affect the conduction mechanism. The black squares with composition x=1 refer to a LSC sample deposited on sapphire single crystal. Data from LSMC bulk sample from the literature are reported for comparison.[35]

All in all, high oxygen diffusion pathways along grain boundaries are found to be on the origin of the excellent mass transport properties presented here for the whole thin film LSMC family. Therefore, grain boundary engineering is presented as a new tool for overcoming a limited performance of bulk materials. This opens a new avenue for exploring new families of interface-dominated pure ionic or mixed ionic-electronic conductors for new applications.

## CONCLUSIONS

A thin film pseudo-binary system of $La_{0.8}Sr_{0.2}(Mn_{1-x}Co_x)_{0.85}O_{3\pm\delta}$ ($x \approx 0$ to 1) has been successfully deposited by combinatorial pulsed laser deposition in order to study the oxygen mass transport properties at the bulk and grain boundary levels as a function of the cobalt content. Isotope exchange depth-profiling coupled with SIMS measurement has been employed to obtain the oxygen self-diffusion ($D_b^*$, $D_{gb}^*$) and surface exchange ($k_b^*$, $k_{gb}^*$) coefficients for a large range of Co compositions at T=600-700ºC ($x = 0$ to 0.8). While the bulk properties follow the behaviour reported in the literature, this study reveals that grain boundaries act as fast diffusion and exchange pathways for oxygen ion transport dominating the transport properties of the whole material in the range of compositions under study. Moreover, supported by the electronic measurements carried out on the whole LSMC family, the high oxygen exchange coefficients observed in Mn-rich compositions seem to be dominated by the high availability of oxygen vacancies, predicted by previous simulation studies. These results extend the striking enhancement, previously reported,[12,14,15] of mass transport properties in interface-dominated electronic conductors (LSM) to good ionic conductors (LSMC), and indicates that fast diffusion pathways found at the grain boundary level are not directly coupled to the bulk performance. The implications of these results are anticipated to be of great relevance for technologies that rely on mixed ionic-electronic conductor materials such as micro-solid oxide fuel cells, lithium ion batteries or nanoionics-based resistive switching devices, particularly if they are used in thin film form.

## ACKNOWLEDGMENT

The research was supported by the Generalitat de Catalunya-AGAUR (2017 SGR 1421), the CERCA Programme and the European Regional Development Funds (ERDF, "FEDER Programa Competitivitat de Catalunya 2007-2013"). This project has received funding from the European Research Council (ERC) under the European Union's Horizon 2020 research and innovation programme (ULTRASOFC, Grant Agreement number: 681146). M.B. would like to thank for the financial support of the Juan de la Cierva postdoctoral programme and A. M. S of the FI predoctoral programmes (EU-European Social Fund).

## ASSOCIATED CONTENT

**Supporting Information.** Details of the experimental and computational methods: Section S.1 - Deposition of combinatorial LSMC films by Pulsed Laser Deposition; Section S.2 - Optimization of the deposition of large-area parent compounds; Section S.3 - Compositional and (micro) structural characterization of the combinatorial sample; Section S.4 - FEM simulations of the isotopic diffusion profiles of the LSMC/YSZ/Si system; Section S.5 - In-plane electrical characterization of LSMC sample.
This material is available free of charge via the Internet at http://pubs.acs.org.

## AUTHOR INFORMATION

### Corresponding Author

* Prof. Albert Tarancón: atarancon@irec.cat

### Author Contributions

The manuscript was written through contributions of all authors. / All authors have given approval to the final version of the manuscript.